\documentclass[aps,prd,12point,twocolumn,nofootinbib,showpacs,superscriptaddress]{revtex4-1}
\def\theequation{\arabic{section}.\arabic{equation}}
\usepackage{psfrag}
\usepackage{subfigure}
\usepackage{color}
\usepackage{mathrsfs}
\usepackage{graphicx}
\usepackage{amssymb, bm}
\usepackage{amsmath, amsthm}
\usepackage{epstopdf}
\usepackage{hyperref}
\usepackage{enumerate}
\usepackage{longtable}

\newcommand{\be}{\begin{equation}}
\newcommand{\ee}{\end{equation}}

\renewcommand{\d}{\mbox{${\rm d}$}}
\newcommand{\rh}{r_{\rm H}}
\newcommand{\Rh}{R_{\rm H}}

\begin{document}
\def\theequation{\arabic{section}.\arabic{equation}} 

\title{Quasilocal mass in scalar-tensor gravity: spherical symmetry}

\author{Andrea Giusti}
\email[]{agiusti@ubishops.ca}
\affiliation{Department of Physics and Astronomy, Bishop's University, 
2600 College Street, Sherbrooke, Qu\'ebec, 
Canada J1M~1Z7}

\author{Valerio Faraoni}
\email[]{vfaraoni@ubishops.ca}
\affiliation{Department of Physics and Astronomy, Bishop's University, 
2600 College Street, Sherbrooke, Qu\'ebec, 
Canada J1M~1Z7}



\begin{abstract}

A recent generalization of the Hawking-Hayward quasilocal energy to 
scalar-tensor gravity is adapted to general spherically symmetric 
geometries. It is then applied to several black hole and other spherical 
solutions of scalar-tensor and $f({\cal R}) $ gravity. The relations of 
this quasilocal energy with the Abreu-Nielsen-Visser gauge and the Kodama 
vector are discussed.

\end{abstract}

\pacs{}

\maketitle

\section{Introduction}
\label{sec:1}
\setcounter{equation}{0}

Einstein's theory of general relativity (GR) predicts spacetime 
singularities where it breaks down and clashes with quantum mechanics in 
the ultraviolet 
regime since it cannot be quantized in any standard way. Therefore, GR is 
expected to be modified at high energy. The 
attempts to quantum-correct GR  produce, in 
the low-energy limit, higher derivative equations 
or extra fields that couple explicitly to the spacetime curvature.  For 
example, the low-energy limit of string 
theories contains a dilaton very similar to the scalar field of 
Brans--Dicke gravity and, in this limit, bosonic string theory 
reduces to a Brans--Dicke theory \cite{bosonic}. 

In the infrared regime, compelling motivation for the study of alternative 
gravity comes from cosmology. The standard model of cosmology, the 
$\Lambda$--Cold Dark Matter ($\Lambda$CDM) model, fits into GR the current 
accelerated expansion of the universe discovered with high redshift 
supernovae only at the price of introducing an extremely fine-tuned 
cosmological constant $\Lambda$ or a completely {\em ad hoc} dark energy 
\cite{AmendolaTsujikawabook}. To avoid invoking either one of those, 
cosmologists consider very seriously the possibility that gravity departs 
from GR at large (cosmological) scales or low densities. The  most 
popular class of 
theories 
studied for this purpose is $f({\cal R})$ gravity \cite{CCT}, which is the 
subject of a large literature \cite{Salvbook, reviews}. This is a subclass 
of scalar-tensor gravity. Scalar-tensor theories \cite{ST}, which 
generalize the original Brans--Dicke theory \cite{BD}, are minimal 
modifications of GR in the sense that they introduce only a scalar degree 
of freedom $\phi$ in addition to the spin two field represented by the 
metric tensor $g_{ab}$ of GR. However, they still exhibit a rich 
phenomenology. Considerable theoretical and experimental effort is 
being put into testing gravity at all scales to either detect or constrain 
deviations from GR in the study of cosmology, black holes, or astrophysics 
\cite{Padilla, Bertietal2013}, including the search for scalar 
hair \cite{scalarhair}.

The (Jordan frame) action of scalar-tensor gravity is\footnote{We follow 
the 
notation of Ref.~\cite{Waldbook} and use units in which Newton's constant 
$G$ and the speed of light $c$ are unity, but sometimes we restore $G$ for 
convenience.}
\begin{eqnarray}
S_{\rm ST} &=& \frac{1}{16\pi} \int \d^4x \sqrt{-g} \left[ \phi {\cal R} 
-\frac{\omega(\phi )}{\phi}  \, \nabla^c\phi \nabla_c\phi -V(\phi) \right] 
\nonumber\\
&&\nonumber\\
&\, & +S^{(m)} \,, \label{STaction}
\end{eqnarray}
where ${\cal R}$ is the Ricci scalar of the metric $g_{ab}$ with 
determinant $g$, the positive Brans-Dicke scalar $\phi $ is approximately 
equivalent to the inverse of the effective gravitational 
coupling strength, 
\be
G_\text{eff}=\phi^{-1} \,,
\ee
$\omega(\phi)$ 
(a constant parameter in the original Brans-Dicke theory 
\cite{BD}) is the ``Brans-Dicke coupling'', and $V(\phi)$ is a 
scalar field potential. $S^{(m)}=\int \d^4x \sqrt{-g} \, {\cal L}^{(m)} $ 
is the  matter action. 

The variation of the action~(\ref{STaction}) with respect to $g^{ab}$ and  
$\phi$ produces the (Jordan frame) field equations \cite{BD, ST} 
\begin{eqnarray}
R_{ab} - \frac{1}{2}\, g_{ab} \mathcal{R} &=& \frac{8\pi}{\phi} \,  T_{ab}^{(m)} 
\nonumber\\
&&\nonumber\\
&\, & + \frac{\omega}{\phi^2} \left( \nabla_a \phi 
\nabla_b \phi -\frac{1}{2} \, g_{ab} 
\nabla_c \phi \nabla^c \phi \right) \nonumber\\
&&\nonumber\\
&\, &  +\frac{1}{\phi} \left( \nabla_a \nabla_b \phi 
- g_{ab} \Box \phi \right) 
-\frac{V}{2\phi}\, 
g_{ab} \,, \nonumber\\
&& \label{BDfe1} \\
\Box \phi = \frac{1}{2\omega+3} & & 
\left( 
\frac{8\pi T^{(m)} }{\phi}   + \phi \, \frac{\d V}{\d \phi} 
-2V -\frac{\d \omega}{\d \phi} \nabla^c \phi \nabla_c \phi \right)  
\nonumber\\
&& \label{BDfe2}
\end{eqnarray}
where $ T^{(m)} \equiv g^{ab}T_{ab}^{(m)} $ is the trace of the matter 
stress-energy tensor  $T_{ab}^{(m)} $.

(Metric) $f({\cal R})$ gravity is described by the action 
\be
S_{f ({\cal R})}= \frac{1}{16\pi} \int \d^4x \sqrt{-g} \, f({\cal R})   
+S^{(m)}  \,,\label{f(R)action}
\ee
where $f({\cal R})$ is a nonlinear function of the Ricci scalar  
and  $S^{(m)}$ is again the matter action.  The action 
$ S_{f({\cal R})}$ is  equivalent to that of a 
Brans-Dicke gravity with Brans-Dicke field $\phi =f'({\cal R})$, 
coupling $\omega=0$, and the rather complicated 
scalar field  potential \cite{reviews} 
\be
V(\phi) ={\cal R} f'({\cal R}) -f({\cal R}) \bigg|_{{\cal R}={\cal 
R}(\phi)} \,, \label{f0}
\ee
where ${\cal R}$ is now a function of the scalar field $\phi=f'({\cal R})$ 
and a  prime 
denotes differentiation with respect to the curvature scalar. 
The relation ${\cal R}= {\cal R} (\phi)$ is not  explicitly invertible in 
general, and the 
potential $V(\phi)$ remains an implicit function of $\phi$.

The field equations are of fourth order,  
\begin{equation}
f'({\cal R}) R_{ab}-\frac{ f({\cal R})}{2} \, g_{ab}=8\pi T_{ab}^{(m)}+  
\nabla_a \nabla_b f'({\cal R}) - g_{ab} 
\square f'({\cal R}) 
\end{equation}
and can be written as the effective Einstein equations \cite{reviews}
\be
R_{ab}-\frac{1}{2} \, g_{ab} {\cal R} = 8\pi  \, \left( 
\frac{T_{ab}^{(m)} }{f'({\cal R}) }+ T_{ab}^{(eff)} \right) 
\,,\label{effectiveEFE}
\ee
where
\begin{eqnarray}
T_{ab}^{(eff)} &=& \frac{1}{8\pi f'({\cal R})} \Big[ \nabla_a \nabla_b 
f'({\cal R}) -g_{ab} \Box f'({\cal R})  \nonumber\\
&&\nonumber\\
&\, &  +\frac{ f({\cal R})-{\cal R}f'({\cal R})}{2} \, g_{ab} \Big] \,.
\end{eqnarray}

In general, minimal requirements on a $f({\cal R})$ theory of gravity are 
that $\phi=f'(R)>0$ in order for the graviton to carry positive 
kinetic energy, and $f''(R)>0$ to avoid the 
notorious Dolgov--Kawasaki instability 
that makes the scalar $\phi$ tachyonic \cite{DolgovKawasaki, mattmodgrav, 
reviews}.

In GR, the concept of mass of a relativistic gravitating system has been 
scrutinized intensely. Gravitational energy cannot be localized as 
a consequence of the equivalence principle and research has turned to 
quasilocal notions, {\em i.e.}, to the energy enclosed by a compact 
spacelike 2-surface.  Several definitions of quasilocal energy have been 
studied, see \cite{Szabados} for a review and 
Ref.~\cite{isolatedhorizons} for a review of the isolated horizon 
formalism containing  more recent energy definitions.  A common feature 
of quasilocal 
energies is that they tend to remain the domain of mathematical physics 
with no practical applications. Recently, we have applied the 
Hawking--Hayward quasilocal construct \cite{HawkingQLE, Hayward, 
Haywardspherical} to cosmology, gravitational lensing, and black holes 
\cite{applications}, and we focus on the Hawking--Hayward quasilocal energy 
here.

The knowledge of quasilocal energy is important in other areas of 
research: it appears in the first law of thermodynamics for gravity. Black 
hole thermodynamics is a well developed theoretical subject, while the 
thermodynamics of gravity and spacetime ({\em e.g.}, \cite{SptThermo}) is 
much more speculative and still under development. In black hole 
thermodynamics, the Hawking-Hayward quasilocal energy is usually assumed 
to be the internal energy of the black hole. Spacetime thermodynamics 
usually extends the range of theories of gravity beyond GR. Since 
alternative gravity is so prominent in all the areas of research 
mentioned, it is essential for their progress to know 
whether the 
quasilocal energy construct extends to these theories, and we begin with 
the simplest and most popular alternative, scalar-tensor gravity 
(Ref.~\cite{Hideki} extends the Hawking--Hayward construct for spherical 
symmetry to $n$--dimensional Lovelock gravity). There are a few 
quasilocal prescriptions in scalar-tensor gravity, and they all 
disagree with each other to some extent \cite{Cai, Zhang, 
Zheng, WuWangYang, 
Cognola, mySTquasilocal, Faycal}. The prescriptions of 
\cite{mySTquasilocal} and \cite{Faycal} agree only {\em in vacuo}; those 
of  Refs.~\cite{Cai, 
Zhang, Zheng, WuWangYang, Cognola} are limited by severe restrictions, 
including  
~$f({\cal  R})$ gravity only; spherical symmetry only; special 
spacetime geometries only, or given only at  black hole horizons. These 
prescriptions are obtained using spacetime 
thermodynamics and the first law \cite{Cai, Zhang, Zheng, WuWangYang, 
Cognola}, 
but there is much uncertainty on the correct thermodynamical quantities to 
use (temperature, entropy, work density, and heat supply vector), which 
reflect in some arbitrariness in any definition of quasilocal energy based 
on the first law.  Moreover, the horizon temperature is a semiclassical 
concept involving difficult calculations in quantum field theory on curved 
spacetime which are hard to complete (thus far, only the tunneling method 
seems to deliver definite results in non-stationary black hole 
geometries). 
The prescription of \cite{mySTquasilocal} is not restricted to $f({\cal 
R})$ gravity nor to special metrics, spherical symmetry, or asymptotic 
flatness and is obtained purely classically and independent of 
thermodynamics by writing the scalar-tensor field equations as effective 
Einstein equations and using the geometric derivation of the 
Hawking--Hayward mass in this ``effective GR'' context.

Here we develop the prescription for a generalization of the 
Hawking--Hayward quasilocal mass to scalar-tensor (including $f({\cal R})$ 
gravity) given in Ref.~\cite{mySTquasilocal}. In view of future 
applications, we provide a general formula for spherical symmetry and we 
apply it to several spherical solutions of scalar-tensor and $f({\cal R})$ 
gravity. As a first test, the new quasilocal mass of \cite{mySTquasilocal} 
reproduces \cite{monopole} the monopole term in the multipole expansion of 
asymptotically flat solutions of scalar-tensor gravity 
\cite{SotiriouPappas}.

\section{Spherical symmetry in scalar-tensor gravity}
\label{sec:2}
\setcounter{equation}{0}

In Einstein's theory, the Hawking-Hayward quasilocal mass is defined  
\cite{HawkingQLE, Hayward} on an embedded spacelike, compact, and 
orientable 
2-surface ${\cal S}$  with induced 2-metric $h_{ab}$ and  induced Ricci 
scalar ${\cal  R}^{(h)}$. Consider ingoing ($-$) and outgoing ($+$) null  
geodesic congruences from  ${\cal S}$ and let $\theta_{(\pm)}$ 
and $\sigma_{ab}^{(\pm)}$ be the expansions and shear 
tensors of these congruences, respectively. Let $\omega^a$ be the 
projection of the commutator of 
the null normal vectors to ${\cal S}$ onto ${\cal S}$  (the 
anoholonomicity \cite{Hayward}). Let  $\mu$ denote the volume 
2-form on  ${\cal S}$ and let  $A$ be the area of ${\cal S}$. Then, the  
Hawking-Hayward quasilocal energy is defined as \cite{HawkingQLE, 
Hayward}
\begin{eqnarray}
M_\text{HH} &=& \frac{1}{8\pi G} \sqrt{ \frac{A}{16\pi}} \int_{\cal S} \mu 
\left(
{\cal R}^{(h)} +\theta_{(+)} \theta_{(-)} -\frac{1}{2} \, \sigma_{ab}^{(+)} 
\sigma^{ab}_{(-)} \right.\nonumber\\
&&\nonumber\\
&\, & \left.  -2\omega_a\omega^a \right) \,. \label{HHmass}
\end{eqnarray}
It can be shown that the quasilocal mass has  a Newtonian character 
because, for an observer with four-velocity parallel to the normal to the 
2-surface ${\cal S}$, only the electric part of the Weyl tensor 
contributes to $M_\text{HH}$ \cite{Symmetryquasilocal}.

The contracted Gauss equation \cite{Hayward}
\be
{\cal R}^{(h)} +\theta_{(+)} \theta_{(-)} -\frac{1}{2} \, \sigma_{ab}^{(+)} 
\sigma^{ab}_{(-)}  = h^{ac}h^{bd} R_{abcd}
\ee
is useful to compute the first three terms in the integral and was used  
in \cite{mySTquasilocal}. 

The scalar-tensor mass prescription of \cite{mySTquasilocal} is 
\begin{eqnarray}
&&M_\text{ST} =  \frac{1}{8\pi} \sqrt{ \frac{A}{16\pi}} 
\int_{{\cal S}}\mu
\phi \left[  h^{ac} h^{bd} C_{abcd} -2\omega_a\omega^a 
\right. \nonumber\\
&&\nonumber\\
& & +\frac{8\pi}{\phi} \, h^{ab}T_{ab}- \frac{16\pi T}{3\phi} 
+\frac{h^{ab}\nabla_a\nabla_b \phi}{\phi} \nonumber\\
&&\nonumber\\
& & \left. +\frac{\omega}{\phi^2} \left( 
h^{ab}\nabla_a \phi \nabla_b \phi 
-\frac{1}{3}\, \nabla^c\phi \nabla_c \phi \right) 
 +\frac{V}{3\phi} \right] \,,\label{generalresult}
\end{eqnarray}
where the $\phi$ factor in the first term on the right hand 
side is introduced by the replacement $G \rightarrow 
 G_\text{eff}$.

In GR, in spherical symmetry, the Hawking--Hayward quasilocal 
energy~(\ref{HHmass}) reduces \cite{Haywardspherical} to the better known 
Misner--Sharp--Hernandez mass \cite{MSH}
\be \label{MSH}
M_\text{MSH}=\frac{R}{2G} \left( 1- \nabla^c R \nabla_c R 
\right) \,,
\ee
where $R$ is the areal radius. Let us consider now  scalar-tensor 
gravity: assuming 
spherical symmetry and the surface ${\cal S}$ to be a 
2-sphere of symmetry with areal radius $R$ and induced metric $h_{ab}$, 
the line element can always be diagonalized as 
\begin{eqnarray}
\d s^2 &=&g_{00}\d t^2 +g_{11} \d R^2 +R^2 \d\Omega_{(2)}^2 \nonumber\\
&&\nonumber\\
&=&  I_{\mu \nu}\d x^\mu \d x^\nu +h_{\mu \nu} \d x^\mu \d x^\nu \label{diagonal}
\end{eqnarray}
in spherical coordinates $\left( t, R, \theta, \varphi \right)$. 
Here $ I_{\mu \nu}= {\rm diag}\left( g_{00}, g_{11}, 0, 0 \right) $,  
$ h_{\mu \nu}={\rm diag}\left(0, 0, R^2, R^2 \sin^2 \theta \right) $, and 
$\d \Omega_{(2)}^2 \equiv \d \theta^2 + \sin^2\theta \d \varphi^2 $ is 
the metric on the unit 2-sphere. Equation (\ref{generalresult})  
then simplifies to \cite{mySTquasilocal}
\begin{eqnarray}
M_\text{ST} &=&  \frac{\phi R^3}{4} 
\left[  h^{ac} h^{bd} C_{abcd} 
+\frac{8\pi}{\phi} \, h^{ab}T_{ab}- \frac{16\pi T}{3\phi} \right. \nonumber\\
&&\nonumber\\
&\, & \left. +\frac{\omega}{\phi^2} \left( h^{ab}\nabla_a \phi\nabla_b \phi 
-\frac{1}{3}\, \nabla^c\phi \nabla_c \phi \right) 
\right.\nonumber\\
&&\nonumber\\
& \,& \left. +\frac{h^{ab}\nabla_a\nabla_b \phi}{\phi}  
+\frac{V}{3\phi} \right] \,.\label{sphericalresult}
\end{eqnarray}

The scalar-tensor quasilocal mass of spheres in 
Friedmann--Lema\^itre--Robertson--Walker (FLRW) spacetimes, given in 
Ref.~\cite{mySTquasilocal}, follows immediately from 
Eq.~(\ref{sphericalresult}). However, here we want to provide a simple 
formula for the scalar-tensor quasilocal mass valid for {\em any} 
spherically symmetric metric and Eq.~(\ref{sphericalresult}) is not the 
most convenient starting point. Let us return instead to the starting 
point used in \cite{mySTquasilocal}) to obtain  
Eq.~(\ref{sphericalresult}), that is, the expression 
\begin{eqnarray}
M_\text{ST} &=& \frac{1}{8\pi } \sqrt{ \frac{A}{16\pi}} \int_{\cal S} \mu \phi  
\left(
{\cal R}^{(h)} +\theta_{(+)} \theta_{(-)} -\frac{1}{2} \, \sigma_{ab}^{(+)} 
\sigma^{ab}_{(-)} \right.\nonumber\\
&&\nonumber\\
&\, & \left.  -2\omega_a\omega^a \right) \,. \label{STmass}
\end{eqnarray}
If ${\cal S}$ is a 2-sphere of areal radius $R$ (denoted by ${\cal S}_R$),  
and assuming 
that the 
scalar field and the metric components in the gauge~(\ref{diagonal}) 
depend only on $t$ and $R$ to respect spherical symmetry, then 
$\phi(t,R)$ can be extracted from the sign of integration and the integral 
reduces to the usual Hawking--Hayward mass, so that
\begin{eqnarray}
M_\text{ST}(t,R) &=& \frac{1}{8\pi } \sqrt{ \frac{A}{16\pi}} \, \phi (t,R) 
\int_{ {\cal S}_R} \mu 
\left( {\cal R}^{(h)} +\theta_{(+)} \theta_{(-)} 
\right.\nonumber\\
&&\nonumber\\
&\, & \left. -\frac{1}{2} \, 
\sigma_{ab}^{(+)}  \sigma^{ab}_{(-)}    -2\omega_a\omega^a \right) 
\nonumber\\
&&\nonumber\\
& =& G \, \phi(t,R) M_\text{HH}(t,R)=  G \phi M_\text{MSH} \,.
\end{eqnarray}
Therefore, the sought for formula for the quasilocal mass in scalar-tensor 
gravity and spherical symmetry is simply
\be
M_{\rm ST}=\frac{\phi R}{2} \left( 1-\nabla^c R \nabla_c R \right) \,. 
\label{formula}
\ee
This expression could {\em a priori} have been guessed by replacing $G$ 
with $ G_\text{eff}=1/\phi$ in the expression of the 
Misner--Sharp--Hernandez mass~(\ref{MSH}). One has
\be
1-\frac{2M_\text{ST}}{\phi R}=\nabla^cR \nabla_c R = g^{RR} 
\label{eq:2.10}
\ee
therefore, in the gauge~(\ref{diagonal}) using the areal radius $R$ as the 
radial coordinate, it is always $g_{RR}= \left( 1-2M_\text{ST}/(\phi R) 
\right)^{-1}$. Moreover, if 
the geometry admits horizons, these are located by the roots of the 
equation $\nabla^c R \nabla_c R$ (in any coordinate system, since this is 
a scalar equation) \cite{MSH, AbreuVisser, NielsenVisser, mylastbook}. It 
follows that, 
on a horizon, it is
\be
R_{\rm H}=\frac{2M_\text{ST}(R_{\rm H})}{\phi( R_{\rm H})} \,\label{R=2M}
\ee
generalizing the well known relation between mass and radius of the 
Schwarzschild horizon. Equation~(\ref{R=2M}) applies to black hole 
horizons, wormhole horizon throats, and cosmological horizons, whether 
they are static or time-dependent ({\em i.e.}, apparent) horizons.

Let us come now to $f({\cal R})$ gravity. In this class of 
theories, the quasilocal mass in spherical symmetry becomes
\be
M_{ f({\cal R})}=\frac{ f'({\cal R}) R}{2} \left( 1-\nabla^c R \nabla_c R 
\right) 
\,. 
\label{formula2}
\ee
As a consequence of the fact that now the effective Brans--Dicke scalar 
$\phi=f'({\cal R})$ multiplies the Misner--Sharp--Hernandez mass known from 
GR, the usual condition $f'({\cal R})> 0$ for the gravitational 
coupling to be positive and the graviton to carry positive kinetic energy 
corresponds to the non-negativity of the quasilocal mass.

\section{Abreu--Nielsen--Visser gauge and Kodama vector} 
\label{sec:3}
\setcounter{equation}{0}

A spherical metric can always be written in a diagonal gauge employing the 
areal radius $R$ as the radial coordinate,  as in 
Eq.~(\ref{diagonal}). We have reached the conclusion, with eq.~(\ref{eq:2.10}), that we can 
write
\be
g_{11}=\left( 1-\frac{2M_\text{ST}}{\phi \, R} 
\right)^{-1} = \left( 1-\frac{2 G M_\text{MSH}}{R} \right)^{-1}\,.
\ee
Nobody forbids to write $g_{00}<0$ as
\be
g_{00}=- \mbox{e}^{-2\Phi} \left(1-\frac{2 G M_\text{MSH}}{R} \right)
\ee
with an appropriate function $\Phi(t, R)$, 
so we can always use the Abreu--Nielsen--Visser 
gauge\footnote{Although we 
use the name Abreu--Nielsen--Visser gauge, this kind of parametrization was 
used before, without name, in the black hole  context 
({\em e.g.}, \cite{Bizon}).} 
 \be
\begin{split}
\d s^2=&-\mbox{e}^{-2\Phi} \left(1-\frac{2 G M_\text{MSH}}{R} \right)\d t^2 \\
&+\left(1-\frac{2 G M_\text{MSH}}{R} \right) ^{-1} \d R^2
+R^2 \d \Omega_{(2)}^2 
\,.
\end{split}
\ee

The Kodama vector is always 
defined geometrically 
in the presence of spherical symmetry and, in this gauge, it is given by
\be
K^a = \frac{1}{\sqrt{ -g_{00} \, g_{11} } } 
\left( \frac{\partial}{\partial t} \right)^a 
=\mbox{e}^{\Phi} \left( \frac{\partial}{\partial t} \right)^a 
\ee
From this vector one can then construct the Kodama 4-current
\be
J^a \equiv G^{ab} K_b \, ,  
\ee
which is a covariantly conserved vector. Indeed, in the 
Abreu--Nielsen--Visser gauge one has 
that
\begin{eqnarray}
J^\mu &=& G^{\mu \nu} K_\nu =  
\mbox{e}^{\Phi} \, G^{\mu}_{\,\,\, 0} \nonumber\\
&&\nonumber\\
&=&  \frac{2 \, G \, \mbox{e}^{\Phi}}{R^2} \left(
- M_\text{MSH}' 
\, , \,   
\dot{M}_\text{MSH}  
\, , \, 0 \, , \, 0   \right) \, ,\nonumber\\
&&  
\end{eqnarray}
from which it follows that
\begin{eqnarray}
\nabla_\mu J^\mu  = \frac{1}{\sqrt{-g}} \, \partial _\mu \left( \sqrt{-g} 
\, J^\mu \right) = 0 \, , 
\end{eqnarray}
 with $\sqrt{-g} = \mbox{e}^{-\Phi} \, R^2 \, \sin \theta$, as in 
\cite{AbreuVisser}.

	A special situation occurs if $\Phi=0$, or 
\begin{equation}\label{Jacobson} 
g_{00} \, g_{11}=-1 \,, 
\ee 
which covers 
many spherically symmetric geometries.\footnote{Early work on this 
class 
of geometries includes Refs.~\cite{BondiKilmister60, French77}.} This 
condition was studied in~Ref.~\cite{Jacobson}, with the result that it is 
equivalent to the requirement that the double projection $R_{ab} \ell^a \ell^b $ 
of the Ricci tensor onto radial null vectors $\ell^a$ vanishes.  
Equivalently, the restriction of the Ricci tensor to the $\left( 
t,R\right)$ subspace is proportional to the restriction of the metric 
$g_{ab}$ to this subspace \cite{Jacobson}. Or, the areal radius $R$ 
constitutes an affine parameter along radial null geodesics 
\cite{Jacobson}. In this case, the Kodama vector is not just parallel, but 
it coincides with the time direction. If, further, the metric is static, 
the Kodama vector is also the timelike Killing vector (while, in general, 
when the latter exists, the former is only parallel to it).\\

\section{FLRW geometry}
\label{sec:5}
\setcounter{equation}{0}

The scalar-tensor quasilocal energy for FLRW universes sourced by  
perfect fluids was derived from Eq.~(\ref{generalresult}) in 
\cite{mySTquasilocal}, but it follows immediately from 
Eq.~(\ref{formula}). Given the importance of the FLRW geometry, we recall 
briefly the results of Ref.~\cite{mySTquasilocal}.  The FLRW line element 
\be 
\d s^2=-\d t^2 +a^2(t) \left( \frac{\d r^2}{1-Kr^2} +r^2 \d \Omega_{(2)}^2 
\right)\,,
\ee
where $K=0,\pm 1$ is the normalized curvature index, is spherically 
symmetric about every spatial point and 
the areal radius is 
$R(t,r)=a(t)r$. The prescription~(\ref{formula}) then gives
\begin{eqnarray}
M_\text{ST}(R) &=& \frac{\phi R^3}{2} \left( H^2 +\frac{K}{a^2} \right) 
\\
&&\nonumber\\
&=&  \frac{H^2R^3 \phi}{2}= \frac{4\pi R^3}{3} \left( \rho 
+\rho_{\phi}  \right) \,, \label{anotherlabel}
\end{eqnarray}
where in the last line the Hamiltonian constraint 
\be\label{Hamconstraint}
H^2 = \frac{8\pi \rho}{3\phi} 
-H\, \frac{\dot{\phi}}{\phi} +\frac{\omega}{6} 
\left( \frac{ \dot{\phi}}{\phi} \right)^2 +\frac{V}{6\phi} 
\equiv \frac{8\pi \left( \rho +\rho_{\phi} \right)}{3\phi}
\ee
was used. In Ref.~\cite{mySTquasilocal}, instead, the 
expression~(\ref{anotherlabel}) was obtained  from the more involved 
Eq.~(\ref{generalresult}).

In metric $f({\cal R}) $ gravity, where $\phi=f'({\cal R})$, 
the Hamiltonian constraint reads  \cite{reviews}
\be
H^2 = \frac{1}{3f'} \left[ 8\pi \rho +\frac{ 
{\cal R}f'-f}{2} 
-3H (f')\,\dot{} \, \right] \,,
\ee
and we obtain \cite{mySTquasilocal} 
\begin{eqnarray} 
M_{f({\cal R})} &=& \frac{H^2R^3\phi}{2} \nonumber\\
&&\nonumber\\
&=& 
\frac{4\pi R^3}{3} \, \rho
+\frac{R^3}{2} \left( \frac{ {\cal R} f'-f}{6} 
- H f'' \dot{{\cal R}}  \right) \label{sphericalgeneral}
\end{eqnarray}
which is, of course, equivalent to Eq.~(\ref{formula2}).

\section{General spherical, static, and asymptotically flat solution of 
Brans-Dicke theory}
\label{sec:6}
\setcounter{equation}{0}

Let us consider the original Brans-Dicke theory with a 
constant coupling parameter $\omega$ and a scalar field $\phi$ without 
mass or potential \cite{BD}. Imposing 
that the solution be static, spherically symmetric, and asymptotically 
flat, Hawking has proved that all black holes reduce to the Schwarzschild 
black hole and the scalar field $\phi $ becomes constant outside the 
Schwarzschild event horizon (the statement is more general, as it includes 
all stationary, asymptotically flat black holes of this theory, which then 
reduce to Kerr \cite{Hawking}). The theorem has been generalized to 
arbitrary scalar-tensor theories in which the scalar field does not have 
singularities or zeros on or outside the horizon, and to scalar field 
potentials with minima that allow states of stable equilibrium for $\phi 
$, the exceptions being physically pathological \cite{Bekenstein, 
circumvent, SotiriouFaraoniPRL}. Then, the spherical, static, 
asymptotically flat black hole solution of scalar-tensor gravity which is 
physically relevant is Schwarzschild with a constant $\phi$ and the 
scalar-tensor quasilocal prescription~(\ref{formula}) trivially reduces to 
the Misner--Sharp--Hernandez mass~(\ref{MSH}).

If $V(\phi) \equiv 0$, the most general static, spherically symmetric, and 
asymptotically flat solution of Brans--Dicke theory that is not  a black 
hole is also known 
\cite{Just, Bronnikov1, DamourFarese, ourAgnese,ourSolar}:
\begin{eqnarray}
\d s^2 &=&  -\mbox{e}^{ (\alpha+\beta)/r } \d t^2
+ \mbox{e}^{ ( \beta-\alpha)/r  }
\left( \frac{ \gamma/r }{ \sinh ( \gamma/r ) } \right)^4 \d r^2  \nonumber\\
&&\nonumber\\
&\, & + \mbox{e}^{ (\beta-\alpha)/r }  
 \left(
\frac{ \gamma/r }{ \sinh( \gamma/r) } \right)^2 r^2 \d \Omega_{(2)}^2 
\,, \label{new1}\\
&&\nonumber\\
\phi (r) &=& \phi_0 \, \mbox{e}^{-\beta/r}  \,,
\;\;\;\;\;\;\; \beta= \frac{\sigma}{\sqrt{|2\omega+3|} }  
\label{new2}
\end{eqnarray}
if $\gamma \neq 0$. Here $\alpha, \beta$, and $\gamma$ are parameters 
satisfying the relations
\be
\beta= \frac{\sigma}{\sqrt{|2\omega+3|}} 
\ee
where $\sigma$ is a scalar charge and 
\be
4\gamma^2= \alpha^2+2\sigma^2
\ee
if $\sigma \neq 0$ (if $\sigma$ vanishes, both $\alpha$ and $\gamma$ 
must vanish, but this cannot be seen in this notation and one needs to 
revert to a form of the metric previously used by Wyman \cite{Wyman}).
This solution is conformal to the 
Fisher--Buchdal--Janis--Newman--Winicour--Wyman solution of general relativity 
with a free scalar field \cite{Fisher, Wyman} and, in a certain coordinate 
chart (of limited 
validity) \cite{ourAgnese}, takes the Campanelli--Lousto form  
\cite{CampanelliLousto}. The electrovacuum generalization was found by 
Bronnikov \cite{Bronnikov1}, while special cases were found 
in~\cite{Bergh} 
for 
$\alpha=\beta$, $\alpha=(2\omega+3)\beta$, and 
$\alpha=-(\omega+1)\beta$. An exhaustive investigation of the general 
solutions of the  Bergmann--Wagoner class of scalar-tensor theories was 
given in~\cite{Bronnikov3}.

If the parameter $\gamma=0$, the Jordan  frame  solution is the 
Brans Class~IV geometry \cite{Brans} 
\begin{eqnarray}
\d s^2 &=& -\mbox{e}^{ -2B/r }\d t^2 + \mbox{e}^{ 2B(C+1)/r }
\left( \d r^2+r^2 \d \Omega_{(2)}^2 \right) \,, \nonumber\\
&&\\
\phi&=& \phi_0 \, \mbox{e}^{-BC/r} \,, 
\end{eqnarray}
where 
\be
B=-\frac{(\alpha+\beta)}{2} \,, \;\;\;\;\;\;\;\;\; 
C=- \, \frac{2\beta}{\alpha+\beta}  \,.\label{newlast}
\ee 

Consider now the solution for $\gamma\neq 0$; the areal radius is 
\be
R(r)= \gamma\, \frac{ \mbox{e}^{\frac{\beta-\alpha}{2r}} 
}{\sinh(\gamma/r)}  \,.\label{arealradius}
\ee
When they exist, apparent horizons are the roots of the  equation
\cite{MSH, AbreuVisser} $\nabla^c R \nabla_c R=0 $. A single root 
describes a black hole horizon, while a double root describes a 
wormhole throat, and no roots means no horizons. The physical nature of 
the solutions (\ref{new1})-(\ref{newlast}) was discussed in 
\cite{ourAgnese}. To 
summarize, for $\gamma\neq 0$ the equation for the apparent 
horizons becomes 
\be 
g^{rr} \left( \frac{\d R}{\d r} \right)^2 = 
\sinh^2(\gamma/r)  \left[  \frac{\alpha -\beta}{2\gamma} + \frac{1}{\tanh 
\left( \gamma/ r \right) }   \right]^2  =0 \,;
\ee
if $(\beta-\alpha)/\gamma >0$ a double root exists, corresponding to a  
wormhole throat at 
\be
\label{eq:5.10}
\rh= \frac{2\gamma}{\ln \left( 
\frac{\beta-\alpha+2\gamma}{\beta-\alpha-2\gamma} \right)} 
= \frac{\gamma}{ \tanh^{-1}\left( \frac{2\gamma}{\beta-\alpha}\right)}
\,.
\ee
If $(\beta -\alpha)/\gamma<0$, instead, there is a naked singularity at 
$R=0$ (the general solution~(\ref{new1}) has a spacetime singularity there 
\cite{ourAgnese}) since, for both signs of  $\gamma $, the Ricci 
scalar 
\be
{\cal R}= \frac{\omega \beta^2}{16\gamma^4} \, \mbox{e}^{\left( 
\alpha-\beta \pm 4\gamma\right)/r} 
\ee
diverges as $r \rightarrow 0$  for $\beta -\alpha<4\gamma$ or for 
$\alpha-\beta > 4\gamma$, 
respectively.
The scalar-tensor quasilocal mass on a sphere of radius $r$ is
\begin{eqnarray}
&& M_\text{ST}(r) = \frac{\gamma \phi_0}{2} \frac{ \mbox{e}^{ 
-\frac{\alpha+\beta}{2r}} }{\sinh \left( \gamma/r\right)} 
\nonumber\\
&&\nonumber\\
& & \, \cdot \left\{ 
1-\sinh^2\left( \frac{\gamma}{r} \right) \left[ 
\frac{\alpha-\beta}{2\gamma} + \frac{1}{\tanh \left( \gamma/r \right) }
\right]^2\right\} \,.
\end{eqnarray}
In the case $(\beta-\alpha)/\gamma >0$, the quasilocal mass evaluated at 
the wormhole throat is

\begin{eqnarray}
\nonumber
M_\text{ST}(\rh)&=& 
\frac{\gamma \phi_0}{2}
\frac{ \mbox{e}^{ 
-\frac{\alpha+\beta}{2 \rh}} }{\sinh \left( \gamma/\rh\right)}\\ 
&=& \frac{\gamma \phi_0}{2}
\frac{ \sqrt{(\beta - \alpha +2\gamma)(\beta - \alpha -2\gamma)}}{2 
\gamma}  \nonumber\\
&&\nonumber\\
&\, & \cdot \left( \frac{\beta - \alpha -2\gamma}{\beta - \alpha 
+2\gamma} 
\right)^{\frac{\alpha + \beta}{4 \gamma}} \nonumber\\
&=& \frac{\phi_0}{4}
\frac{ \left(\beta-\alpha-2\gamma\right)^{  
\frac{\alpha+\beta+2\gamma}{4\gamma} } }{
\left(\beta-\alpha+2\gamma\right)^{ 
\frac{\alpha+\beta-2\gamma}{4\gamma}} }     
\,,
\end{eqnarray}
where in the middle line we used \eqref{eq:5.10} and the identity 
$$
\sinh (x) = \frac{\mbox{e}^{2x} - 1}{2 \mbox{e}^{x}} \, ,
$$
which implies
$$
 \sinh (\gamma/\rh) = \frac{2 \gamma}{\sqrt{(\beta - \alpha 
+2\gamma)(\beta - \alpha -2\gamma)}} \, .
$$

For the general spherical, static, and asymptotically flat  solution of 
Brans--Dicke gravity,  the scalar-tensor mass~(\ref{formula}) reproduces 
\cite{monopole} the monopole term found in a multipole expansion of the 
scalar-tensor metric \cite{SotiriouPappas}.

Let us examine now the $\gamma=0$ case giving the Brans Class~IV solution. 
The areal radius is 
\be
R(r) = \mbox{e}^{ \frac{B(C+1)}{r}} r
\ee
and the equation 
locating the apparent horizons reduces to 
\be
\nabla^c R \nabla_c R= 
\left[ 1-\frac{B(C+1)}{r} \right]^2=0 \, ,
\ee 
which has  a double root $\rh=B(C+1)=(\beta-\alpha)/2$ corresponding to a  
wormhole throat if $\beta>\alpha$ and to a central naked singularity 
otherwise \cite{VFS, ourAgnese}. The quasilocal mass~(\ref{formula}) in a  
sphere of radius $r$ is
\begin{eqnarray}
M_{\rm ST}(r) 
&=& \frac{\phi_0 \, r}{2} \mbox{e}^{B/r} \left\{
1 - \left[ 1-\frac{B(C+1)}{r} \right]^2
\right\}  \nonumber \\
&& \nonumber\\
 &=& \phi_0 \, \mbox{e}^{B/r} \rh \left(1-\frac{\rh}{2r}\right) \,. 
\end{eqnarray}
In the case of the naked central singularity, $M_{\rm ST}(r)$ is negative in 
the central region $0<r< \rh/2$ (as is common for naked singularities, for 
example for the Schwarzschild solution of GR with negative mass) and 
positive for $r>\rh/2$.
When there is a wormhole throat ({\em i.e.}, for $\beta>\alpha$, the 
quasilocal mass on the throat is
\be
M_{\rm ST}(\rh) = \frac{\phi_0 \, \rh}{2} \, \mbox{e}^{B/\rh}
= \frac{\phi_0 \, (\beta - \alpha)}{4} \, \mbox{e}^{\frac{\alpha+\beta}{\alpha-\beta}}
\ee
and it is positive.

\section{Static geometry but time--dependent mass}
\label{sec:7}
\setcounter{equation}{0}

Situations can arise in which the geometry is static but the quasilocal 
mass is time-dependent because $\phi$ is not static. As an example 
consider the special solution of Brans--Dicke theory with $\omega=-1$ 
(the value of the Brans--Dicke parameter corresponding to the low-energy 
limit of bosonic string theory \cite{bosonic}) and 
linear 
potential $V(\phi)=V_0 \phi$ found in 
\cite{confonarev, DilekShawn} in the Campanelli--Lousto 
\cite{CampanelliLousto} form
\begin{eqnarray}
\d s^2 &=& -\d t^2 +A(r)^{-\sqrt{2}} \d r^2 + A(r)^{1-\sqrt{2}} r^2 
\d \Omega_{(2)}^2 \,, \label{special1} \nonumber\\
&&\\
\phi(t,r) &=& \phi_0 \, \mbox{e}^{2at} A(r)^{1/\sqrt{2}} 
\,,\label{special2}
\end{eqnarray}
where $A(r) =1-2m/r$ and $a$ and $m$ are parameters. 

The areal radius is
\be
R(r)= \left(1-\frac{2m}{r} \right)^{\frac{1-\sqrt{2}}{2} } r 
\ee
and the equation locating the apparent horizons  is 
\cite{confonarev}
\begin{eqnarray}
\nabla^c R \nabla_c R &=& g^{rr} \left( \frac{\d R}{\d r} \right)^2 
\nonumber\\
&&\nonumber\\
&=& 
A(r)^{-1} 
\left[ 1-\left( 1+\sqrt{2} \right)\frac{m}{r} \right]^2 =0 \,.
\end{eqnarray}
For $m>0$ there is always a double root, corresponding to a  wormhole 
throat at $\rh=(1+\sqrt{2})m$ or proper radius 
 $\Rh=  \left(1+\sqrt{2} \right)^{\sqrt{2}} m  \simeq 3.48 
\, m$. The quasilocal mass at this throat is
\be
M_{\rm ST}(\Rh) = \frac{\phi(\Rh) \Rh}{2} = 
\frac{m\, \phi_0 \, \mbox{e}^{2at} }{2} 
\,.
\ee
Naively, one would expect the ``mass'' to be $m $ and to be constant but, 
although the wormhole 
throat at $\Rh$ does not change in time, the quasilocal mass 
$M_\text{ST}(\Rh)$ depends on time through $\phi(\Rh)$.

The situation in which the scalar field does not share the symmetries of 
the spacetime geometry is known to generate stealth solutions and violate 
the no-hair theorems in Horndeski and generalized Horndeski theories. One 
possibility is to introduce a linearly time-dependent scalar field profile 
\cite{circumvent, stealth}. If such a solution is found in the more 
conventional 
scalar-tensor theory~(\ref{STaction}), then, through $\phi(t)$, the 
quasilocal mass~(\ref{formula}) will be time-dependent even though the 
geometry is stationary.

\section{The BBMB maverick solution for conformal coupling}
\label{sec:8}
\setcounter{equation}{0}

Nonminimal coupling to the Ricci scalar ${\cal R}$   
appears when a canonical, minimally coupled test 
scalar field $\psi$ is quantized on a curved space \cite{CCJ1} and also, 
classically, in the  context of 
radiation problems (\cite{ChernikovTagirov, DeWittBrehme, 
SonegoFaraoni}, see also \cite{Odintsov5, CCJ2, CCJ3, CCJ4, Friedlander}). 
The nonminimal 
coupling of the scalar $\psi$ has been studied extensively in 
early universe inflation  (\cite{NMCinflation} and references therein).
When the scalar is allowed to 
gravitate, one has a scalar-tensor theory 
\cite{FujiiMaeda, mySTbook, Salvbook} with action 
\begin{eqnarray}
S_\text{NMC} &=& \int \d^4x \sqrt{-g} \Big[ \left( \frac{1}{8\pi  G} -  \xi  
\psi^2 \right) \frac{{\cal R}}{2} - \frac{1}{2}\, \nabla^e \psi 
\nabla_e \psi  \nonumber\\
&&\nonumber\\
&\, &  - V(\psi) \Big] \,,\label{NMCaction}
\end{eqnarray}
where $\xi$ is the dimensionless coupling constant (with $\xi=1/6$ 
corresponding to conformal coupling \cite{CCJ1, Waldbook}), the value of 
which depends on the nature of the scalar and can often be determined 
as a running coupling going to an infrared fixed point under a 
renormalization group 
flow \cite{xi-value, Odintsov5}.  The general scalar-tensor action is 
given by Eq.~(\ref{STaction}) instead of (\ref{NMCaction}),  but  it is 
sufficient to write 
\begin{equation}
    \phi = \frac{1 - 8\pi G \xi\psi^2}{G}  \,,\label{bohphi}
\end{equation}
and use
\begin{eqnarray}
\psi & = & \pm \sqrt{ \frac{1-  G\phi}{ 8\pi  G \xi}}  \,, \\
&&\nonumber\\
\nabla_e \psi & = & \mp  \sqrt{\frac{G}{32\pi \xi 
\left(1 - G\phi\right)}} \, \nabla_e \phi  \,, 
\end{eqnarray}
to reduce~(\ref{NMCaction}) to the standard form~(\ref{STaction}) with
\be
 \omega(\phi)  = \frac{G\phi}{4 \xi \left(1 - G\phi \right) } 
\ee  
Contrary to the Brans-Dicke field $\phi$, the nonminimally coupled  
scalar $\psi$ is not restricted to be positive. However,  
for $\xi>0$ the scalar $\psi$ must satisfy $|\psi|< 
\psi_c \equiv 1/\sqrt{ 8\pi G \xi}$, while all values of $\psi$ are 
admissible if $\xi<0$. 

The Bocharova--Bronnikov--Melnikov--Bekenstein (BBMB) solution of 
conformally coupled ($\xi=1/6$) Einstein-scalar field theory found 
in~\cite{BBM70} was rediscovered in \cite{Bekenstein74}.  This is 
a black hole solution with event horizon and scalar hair, but the scalar 
field $\psi$ is singular on the horizon. This property is unphysical 
\cite{Bekenstein74}, making this solution a maverick. The BBMB solution is 
also unstable with respect to linear perturbations 
\cite{BronnikovKireyev78}.

Following the derivations of \cite{BBM70} and \cite{Bekenstein74}, 
Xanthopoulos and Zannias \cite{XanthopoulosZannias91} and Klimc\'{i}k 
\cite{Klimcik93} proved explicitly that the BBMB construct is the unique 
solution of the Einstein-conformal scalar field equations which is static, 
spherical, asymptotically flat, and does not have constant $\psi$. A new 
proof of the uniqueness of the BBMBM solution outside the photon surface 
(the surface composed of the unstable circular photon orbits) was given 
recently in Ref. \cite{TomikawaShiromizuIzumi17}.  The BBMB solution has 
also been 
generalized by including a cosmological constant, a quartic potential 
$V(\psi)=\lambda \psi^4$, a Maxwell field, different horizon topologies 
\cite{MartinezTroncosoZanelli03, VirbhadraParikh94a, 
VirbhadraParikh94b,MartinezStaforelliTroncoso06}, or an accelerating BBMB 
black hole \cite{CharmousisKolyvarisPapantonopoulos09}.

In the Abreu--Nielsen--Visser gauge it is $\Phi=0$ and the  BBMB 
solution reads \cite{Bekenstein74}
\begin{eqnarray}
\d s^2 &=&- \left(1-\frac{m}{R} \right)^2 \d t^2 
+ \frac{\d R^2}{ \left(1-m/R \right)^2} +R^2 \d \Omega_{(2)}^2 \,,\nonumber\\
&&\\
\psi(R) &=& \sqrt{ \frac{3}{4\pi G}} \, \frac{m}{R-m} \,,
\end{eqnarray}
which represents an extremal Reissner-Nordstr\"om black hole with horizon 
at $R=m$, but the scalar field $\psi$ is singular there. Correspondingly, 
the Jordan frame Brans-Dicke-like field given by Eq.~(\ref{bohphi}) 
is negative and divergent at $R=m$. The scalar-tensor quasilocal mass  
of a sphere of radius $R$
is then
\begin{eqnarray}
\nonumber 
M_{\rm ST}(R) &=&
\frac{\phi R}{2} \left( 1-\nabla^c R \nabla_c R \right) \\
&=& 
\nonumber 
\frac{R^2}{2 G}
\frac{R-2m}{(R-m)^2} 
\left[ 1- \left(1-\frac{m}{R} \right)^2 \right] \, ,
\end{eqnarray}
from which one finds
\be
\lim _{R \to m} M_{\rm ST}(R) = - \infty \, .
\ee
The pathology of the scalar field at the horizon 
(divergent $\psi$ or negative and divergent $\phi$, which means vanishing 
gravitational coupling strength) is reflected in this unphysical property of the quasilocal 
mass. Gravity is repulsive, and $M_{\rm ST}(R)$  is negative, in 
the entire region $m < R <2m$ surrounding the horizon, diverging at $R=m$.

\section{Black holes in $f(R)$ gravity}
\label{sec:9}
\setcounter{equation}{0}

Finally, let us examine a class of static, spherically symmetric, and 
asymptotically flat black holes found recently in $f({\cal 
R})= {\cal R}+2\beta \sqrt{ {\cal R}}$ gravity \cite{Elizaldeetal20}. 
In the Abreu-Nielsen-Visser gauge it is again $\Phi=0$ and the line 
element reads 
\be
\d s^2 = -w(R)\d t^2 + \frac{\d R^2}{w(R)} +R^2 \d\Omega_{(2)}^2 \,,
\ee
where 
\be 
w(R)=\frac{1}{2} +\frac{1}{3\beta R} +\frac{\kappa^2}{R^2} \,,
\ee
$ \kappa^2= Q_E^2 +Q_M^2$, $Q_E$ and $Q_M$ are electric and magnetic 
charges, respectively, and $\beta$ is a parameter with the dimensions of 
a mass. For $Q_E=Q_M=0$, the 
solution reduces to an uncharged one found in 
Ref.~\cite{SebastianiZerbini11}. 
The Ricci scalar is ${\cal R}=1/R^2$ \cite{Elizaldeetal20} and the 
Kodama vector coincides with the timelike Killing vector.

Requiring the gravitational coupling to be positive and the theory to be 
locally stable with respect to the Dolgov--Kawasaki (tachyonic) 
instability \cite{DolgovKawasaki, mattmodgrav} implies
\begin{eqnarray}
f'( {\cal R}) &=& 1+\frac{\beta}{\sqrt{{\cal R}}}=1+\beta R >0 \,,\\
&&\nonumber\\
f''( {\cal R}) &=& -\frac{\beta}{2 {\cal R}^{3/2} }=-\frac{\beta R^3}{2}>0 
\,,
\end{eqnarray}
which imply that $\beta<0$ and $R\leq 1/|\beta|$ (therefore, this solution 
can 
only be used as a model in this region). 

The quasilocal mass~(\ref{formula2}) is
\be
M_{f({\cal R})}= \frac{R\left(1-|\beta|R\right)}{4} \left( 
1+\frac{2}{3|\beta|R} -\frac{2\kappa^2}{R^2} \right) \,.
\ee
At the horizons (when they exist), it is
\be
M_{f({\cal R})} (\Rh)= \frac{\Rh}{2} \left( 1- |\beta|\Rh \right) \,.
\ee
Horizons correspond to the roots of $w(R)=0$, therefore 
\cite{Elizaldeetal20}:

\begin{itemize}

\item If $-\frac{1}{3\sqrt{2}}<\kappa \beta < \frac{1}{3\sqrt{2}}$ there 
are two (inner 
and outer) horizons at
\be
R_{\pm}= \frac{1}{3|\beta|} \left( 1\pm \sqrt{1-18 \kappa^2 
\beta^2}\right) \,,
\ee
with 
\be
0<R_{-}<R_{+}<\frac{2}{3|\beta|}< \frac{1}{|\beta|} \,.
\ee
The scalar-tensor mass~(\ref{formula2}) on the outer horizon is
\be
M_{f({\cal R})} (R_{+})= \frac{\left(1+18\kappa^2\beta^2 
+\sqrt{1-18\kappa^2 \beta^2} \right)}{18|\beta|}  
\ee
and is positive. For comparison, the quasilocal mass of \cite{Cognola, 
Zheng} is \cite{Elizaldeetal20}
\be
\bar{M}(R_{+})= \frac{\left(1+9\kappa^2\beta^2 
+\sqrt{1-18\kappa^2 \beta^2} \right) }{12|\beta|} \,.
\ee

\item If $\kappa \beta=\pm \frac{1}{3\sqrt{2}}$ there is a double root 
corresponding to a wormhole throat at $\Rh=(3|\beta|)^{-1}=\sqrt{2}\, 
|\kappa|$. The quasilocal mass at this throat
\be
M_{f({\cal R})} (\Rh)= \frac{1}{9|\beta|}= \frac{\sqrt{2} \, |\kappa| }{3} 
\ee
is also positive.

\item If $\kappa \beta < -\frac{1}{3\sqrt{2}}$ or $\kappa 
\beta>\frac{1}{3\sqrt{2}}$ there are no real roots of 
$\nabla^cR\nabla_c R=0$ and the geometry contains a naked singularity at 
$R=0$, where the Ricci scalar ${\cal R}=1/R^2$ diverges. The quasilocal 
mass of a sphere of radius $R$ is
\begin{eqnarray}
M_{f({\cal R})} (R)&=& \frac{R\left(1-|\beta|R\right)}{2} \left( 
\frac{1}{2} +\frac{1}{3|\beta|R} -\frac{\kappa^2}{R^2} \right) 
\,.\nonumber\\
&&
\end{eqnarray}
In the region $R<1/|\beta|$, this mass is negative when  $3|\beta| R^2 +2 
R - 6 |\beta| \kappa^2 < 0$, which corresponds to $R_1<R<R_2$, where 
$$
R_{1,2}= - \frac{1 \pm \sqrt{1+18 \kappa^2 \beta^2}}{3|\beta|} \, .
$$
Limiting ourselves to the physical region $R>0$, the quasilocal mass 
is negative in the region 
\be
0<R< R_2 =
\frac{\sqrt{1+18 \kappa^2 \beta^2}-1}{3|\beta|} 
\ee
surrounding the naked singularity. Note that $R_2$ can potentially exceed 
$R=1/|\beta|$, in which case the quasilocal mass is negative 
everywhere.

\end{itemize}

\section{Conclusions}
\label{sec:10}
\setcounter{equation}{0}

There is little doubt that the mass-energy of a system is one of the most basic concepts in 
physics and astrophysics, yet GR is ambiguous in this regard, offering several different 
quasilocal energy prescriptions \cite{Szabados}. Moreover, the concept of quasilocal energy 
seems to have remained confined to the realm of formal mathematical physics while, to be useful, 
it should become part of relativistic astrophysics and cosmology. The application of the 
Hawking--Hayward quasilocal prescription \cite{HawkingQLE, Hayward, Haywardspherical} to 
cosmology and astrophysics has been started in \cite{applications}. Since there is currently 
much motivation, especially from cosmology, to explore alternative theories of gravity 
theoretically and observationally, it is useful to extend the quasilocal energy construct of 
\cite{HawkingQLE, Hayward, Haywardspherical} to the prototypical alternative to GR, 
scalar-tensor gravity (which includes the subclass of $f({\cal R})$ theories nowadays very 
popular in cosmology \cite{CCT, reviews}). The most straightforward prescription of quasilocal 
energy in these theories (in the sense that it is based simply on writing the field equations as 
effective Einstein equations and is independent of thermodynamics of spacetime, black hole 
thermodynamics, and the subsequent restriction to black hole horizons) was given recently in 
\cite{mySTquasilocal}. In this work we have discussed this prescription for spherically 
symmetric geometries, which are the simplest situations occurring in the modelling of systems of 
interest in astrophysics and cosmology. The particularly convenient Abreu--Nielsen--Visser metric 
gauge has been discussed, together with its relation with the Kodama vector used in black hole 
thermodynamics. As the case of FLRW cosmology shows, it is much more convenient to derive the 
quasilocal mass of spherical systems from the simple formula~(\ref{formula}) than from the 
general prescription~(\ref{generalresult}). These developments will be be used in future work 
related to black hole thermodynamics and astrophysics in scalar-tensor and $f({\cal R})$ 
gravity. 

\begin{acknowledgments} 

This work is supported, in part, by the Natural Sciences and Engineering 
Research Council of Canada (Grant No.~2016-03803 to V.F.). The work of A.G.
has also been carried out in the framework of activities of the National Group of Mathematical
Physics (GNFM, INdAM).

\end{acknowledgments}


\end{document}